\documentclass[10pt,apjl]{emulateapj}

\usepackage{apjfonts}

\newcommand{\iso}[2]{\hbox{${}^{#1}{\rm #2}$}}
\newcommand{\Msun}{\ensuremath{{\rm M}_{\sun}}}

\shorttitle{Ge production in AGB stars}
\shortauthors{Karakas et al.}

\begin{document}


\title{Germanium Production in Asymptotic Giant Branch stars: \\
  Implications for Observations of Planetary Nebulae}


\author{Amanda I. Karakas\altaffilmark{1,2}}
\affil{Origins Institute and Department of Physics \& Astronomy, McMaster 
     University, \\ 1280 Main Street W., Hamilton ON L8S 4M1, Canada}
\email{karakas@physics.mcmaster.ca}

\author{Maria Lugaro\altaffilmark{2}}
\affil{Sterrenkundig Institute, University of Utrecht, Postbus 80000, 
 3508 TA Utrecht, the Netherlands}
\email{M.Lugaro@phys.uu.nl}

\and

\author{Roberto Gallino\altaffilmark{2}}
\affil{Dipartimento di Fisica Generale, Universit\'a di Torino, Via P. Giuria 1,
        10125 Torino, Italy}
\email{gallino@ph.unito.it}


\altaffiltext{1}{Present address: Research School of Astronomy \& Astrophysics, 
Mount Stromlo Observatory, Cotter Road Weston Creek, ACT 2611, Australia}
\altaffiltext{2}{Centre for Stellar \& Planetary Astrophysics, Monash University,
Clayton VIC 3800, Australia}


\begin{abstract}
Observations of planetary nebulae (PNe) by Sterling, Dinerstein and Bowers have
revealed abundances in the neutron-capture element Germanium (Ge) from solar to
factors of 3 -- 10 above solar.  The enhanced Ge is an indication that the
$slow$-neutron capture process ($s$ process) operated in the parent star during the
thermally-pulsing asymptotic giant branch (TP-AGB) phase.
We compute the detailed nucleosynthesis of a series of AGB models to
estimate the surface enrichment of Ge near the end of the AGB. A partial mixing zone 
of constant mass is included at the deepest extent of each dredge-up episode, resulting
in the formation of a \iso{13}C pocket in the top $\sim 1/10^{\rm th}$ of the He-rich 
intershell. All of the models show surface increases of [Ge/Fe] $\lesssim 0.5$, 
except the 2.5$\Msun$, $Z=0.004$ case which produced a factor of 6 enhancement of Ge. 
Near the tip of the TP-AGB, a couple of extra TPs could occur to account for 
the composition of the most Ge-enriched PNe. 
Uncertainties in the theoretical modeling of AGB stellar evolution
might account for larger Ge enhancements than we predict here.
Alternatively, a possible solution could be provided by the occurrence of 
a late TP during the post-AGB phase. Difficulties related to spectroscopic
abundance estimates also need to be taken into consideration.
Further study is required to better assess how the model uncertainties 
affect the predictions and, consequently, if a late TP should be invoked.
\end{abstract}




\section{Introduction}

After the TP-AGB phase is terminated, low to intermediate mass stars 
(with initial masses $\sim 0.8$ to 8$\Msun$) evolve through the PNe phase before 
ending their lives as white dwarfs 
\citep[see the recent review by][ and references therein]{herwig05}.
The whole envelope is lost by low-velocity stellar winds during the AGB phase.
The gaseous nebula is the remnant of the deep convective envelope that once 
surrounded the core, which is now exposed as the central star (CS) of the illuminated
nebula.  Thus the abundances of the nebula should reveal information about the 
chemical processing that took place during the AGB; and more precisely, 
information about the last TPs.  The spectra of PNe are usually a 
composite, a mixture from the nebula and the illuminating central star.

Accurate abundances from PNe for helium, C, N, O, Ne, S, Cl and Ar
are available \citep{kaler78,henry89,dopita97,stanghellini00}.
These abundances can be used as a powerful tool to constrain models of
AGB stars \citep{karakas03a}, in particular the amount of mixing that
occurs during third dredge-up (TDU) episodes following a TP.
The discovery in PNe of lines of the light neutron-capture
element Ge by \citet*{sterling02} provides not only constraints 
on the amount of dredge-up but also on the operation 
of the $s$ process occurring in AGB stars.  Despite the solar Ge 
abundance being dominated by massive-star nucleosynthesis,
with AGB stars only contributing 6\% of solar Ge \citep{arlandini99,busso01}, 
this element can be produced by the $s$-process in sufficient quantities
to be observed in PNe. 

In this Letter we seek to address the question of whether 
detailed models of AGB stars can reproduce the observed amount of Ge in PNe. 
This study is timely given the increasing number of detections of 
heavy elements in PNe and indeed this is the first study dedicated to comparing 
to these abundances. In a follow-up paper we will expand upon this work to cover 
all elements up to the first $s$-peak. 

\section{The Numerical Method}

The numerical method we use has been previously described in detail
\citep{lugaro04,karakas06a}. Here we summarize the main points relevant for
this study. We compute the structure first and then perform detailed 
nucleosynthesis calculations.  The reaction network in the post-processing 
algorithm has been expanded from 74 species to 156, to include all stable 
isotopes up to \iso{75}As.  
We include all $\beta$-decays up to arsenic, and all proton, 
$\alpha$ and $n$ capture reactions on all species in the network, which 
results in a total of 1260 reaction rates.

To account for neutron captures on elements heavier than arsenic 
we use a ``double neutron-sink'' description \citep*{herwig03a},
and include two artificial species which are linked by the following reactions
\iso{75}As($n, \gamma$)\iso{76}g and \iso{76}g($n,L$)\iso{76}g, where \iso{76}g
replaces \iso{76}As, and has an initial abundance equal to
the sum of solar abundances from Se to Bi.  The second artificial particle,
$L$, is equivalent to counting the number of neutrons captured beyond arsenic.
The ratio ($L$/\iso{76}g) is a description of the neutrons captured per
seed nuclei and could in principle be related to the $s$-process
distribution.

Most of the 1260 reaction rates are from the 1991 updated REACLIB Data Tables
\citep*{thielemann86}. Many of the proton, $\alpha$ and 
$n$ capture-reaction rates have been updated by \citet{lugaro04}.
We use the NACRE \citep{angulo99} rates for the NeNa and MgAl
chains, and rates from \citet{karakas06a} for the \iso{22}Ne $+ \alpha$ reactions.
The cross-section of the \iso{76}g($n,L$)\iso{76}g reaction is a composite
calculated using heavy element distributions produced by the $s$ process 
in low-mass AGB stars from the models of \citet{gallino98}. 
Test simulations show that using a different cross-section for \iso{76}g does 
not significantly effect the elemental abundance of Ge (or any other species) 
in a 3$\Msun$, $Z=0.012$ or a 2.5$\Msun$, $Z = 0.004$ model.

\section{The stellar models}

\begin{table}
\begin{center}
\caption{Details of the stellar models.\label{models}}
\vspace{1mm}
\begin{tabular}{@{}ccccccccc@{}}  \hline \hline
 Mass &  $Z$  &  TPs & $T_{\rm He}^{\rm max}$ & Mass$_{\rm dred}$ 
	& HBB? & C/O & \iso{12}C/\iso{13}C & $M_{\rm env}$ \\ \hline
 3.0  &  0.02  &  26 & 302 & 8.11($-2$) & No  & 1.40 & 118.0 & 0.676 \\
 5.0  &  0.02  &  24 & 352 & 5.03($-2$) & Yes & 0.77 & 7.840 & 1.500 \\
 6.5  &  0.02  &  40 & 368 & 4.70($-2$) & Yes & 0.40 & 11.60 & 1.507 \\
 3.0  &  0.012 &  22 & 307 & 9.09($-2$) & No  & 2.79 & 185.0 & 0.805 \\
 6.5  &  0.012 &  51 & 369 & 6.48($-2$) & Yes & 0.76 & 10.40 & 1.389 \\
 1.9  &  0.008 &  19 & 278 & 1.70($-2$) & No  & 2.08 & 143.0 & 0.221 \\
 2.5  &  0.004 &  30 & 308 & 8.17($-1$) & No  & 14.7 & 1416 & 0.685 \\
\end{tabular}
\end{center}
\end{table}		

We compute stellar models covering a range of initial mass and 
metallicity listed in the first two columns
of Table~\ref{models}. The metallicities were chosen to reflect the
composition of the progenitor AGB stars which were estimated to 
range from $0.3 Z_{\odot}$ to $Z_{\odot}$ \citep{sterling05c}.
The $Z=0.012$ stellar models were computed with the revised solar 
elemental abundances from \citet*{asplund05} for comparison to the
$Z = 0.02$ models with abundances from \citet{anders89}. 
Scaled-solar abundances were assumed for the 
$Z=0.008$ and 0.004 models. We use \citet{vw93} mass loss on the 
AGB.

In Table~\ref{models} we include the number of TPs, the maximum 
temperature in the He-shell, the total mass dredged into the envelope 
during the AGB, the final surface C/O and \iso{12}C/\iso{13}C ratios,
and envelope mass at the last computed time-step.
All masses are in solar units and temperatures in $10^{6}\,$K.
From Table~\ref{models} we see that the 3 and 6.5$\Msun$, $Z=0.012$
models were similar to their $Z=0.02$ counterparts.
The low-mass AGB models became carbon stars, whereas those models
with hot bottom burning (HBB) retained an O-rich atmospheric 
composition (5, 6.5$\Msun$). 
There is evidence \citep[e.g.][]{abia97} that some extra-mixing 
takes place in $\sim \,$1 to 3$\Msun$ AGB stars, resulting in 
lower observed \iso{12}C/\iso{13}C and C/O ratios than predicted 
by models such as ours.

The results presented in this study are subject to many model
uncertainties including convection and mass loss, which affect both
the structure and nucleosynthesis. We refer to \citet{goriely00}, 
\citet{herwig05}, \citet{ventura05a}, and \citet{straniero06}
for detailed discussions on this topic.

\section{The inclusion of a partial-mixing zone}

The inclusion of a partial mixing zone (PMZ) at the deepest extent
of dredge-up will mix protons from the envelope into the He-intershell,
producing a \iso{13}C pocket. In the PMZ neutrons are liberated during the
interpulse period by the reaction \iso{13}C($\alpha,n$)\iso{16}O, and 
are captured by Fe-seed nuclei to produce heavy elements.
Observational and theoretical evidence suggests this is the dominant
neutron source in low-mass AGB stars \citep{smith87,gallino98}. 
The details of how the pocket forms and its extent in mass
in the He-intershell are still unknown although gravity waves \citep{denissenkov03},
convective overshoot \citep{herwig00}, induced overshoot during the TDU 
episodes (Cristallo et al. 2004; Straniero et al. 2006), and 
rotationally-induced mixing \citep{langer99,herwig03a} have been suggested. 

The \iso{22}Ne($\alpha, n$)\iso{25}Mg reaction produces a brief
strong burst of neutrons in the convective pocket during a TP
when the temperature exceeds $\sim 300 \times 10^{6}$K.
This is probably the dominant neutron source in massive AGB
stars, where the mass of the He-shell is smaller by about an order
of magnitude compared to lower-mass stars, reducing the importance of
the \iso{13}C pocket.

As done in previous nucleosynthesis studies \citep{gallino98,goriely00,lugaro04}
we artificially include a PMZ of constant mass. In Table~\ref{results} 
we show the maximum mass of the pulse-driven convective region, 
$M_{\rm csh}$, at the last TP along with the PMZ masses used. 
Note that $M_{\rm csh}$ is approximately equal to the
He-intershell mass during the TP.
The PMZ mass is estimated to be between $\sim 5$\% to 15\% of the mass of the
He-rich intershell, and we used PMZs of $1 \times 10^{-3}$
and $3 \times 10^{-3}\Msun$ for the 3$\Msun$ models.
We chose a proton profile in which the number of protons decreases 
exponentially with the mass depth below the base of the convective 
envelope in the same way as described in \citet{lugaro04}.
We also computed models with {\em no} PMZ for all 
masses, in order to single out the effect of the \iso{22}Ne source.

\section{Results}

\begin{figure}
\begin{center}
\begin{tabular}{c}
\includegraphics[width=7cm,angle=90]{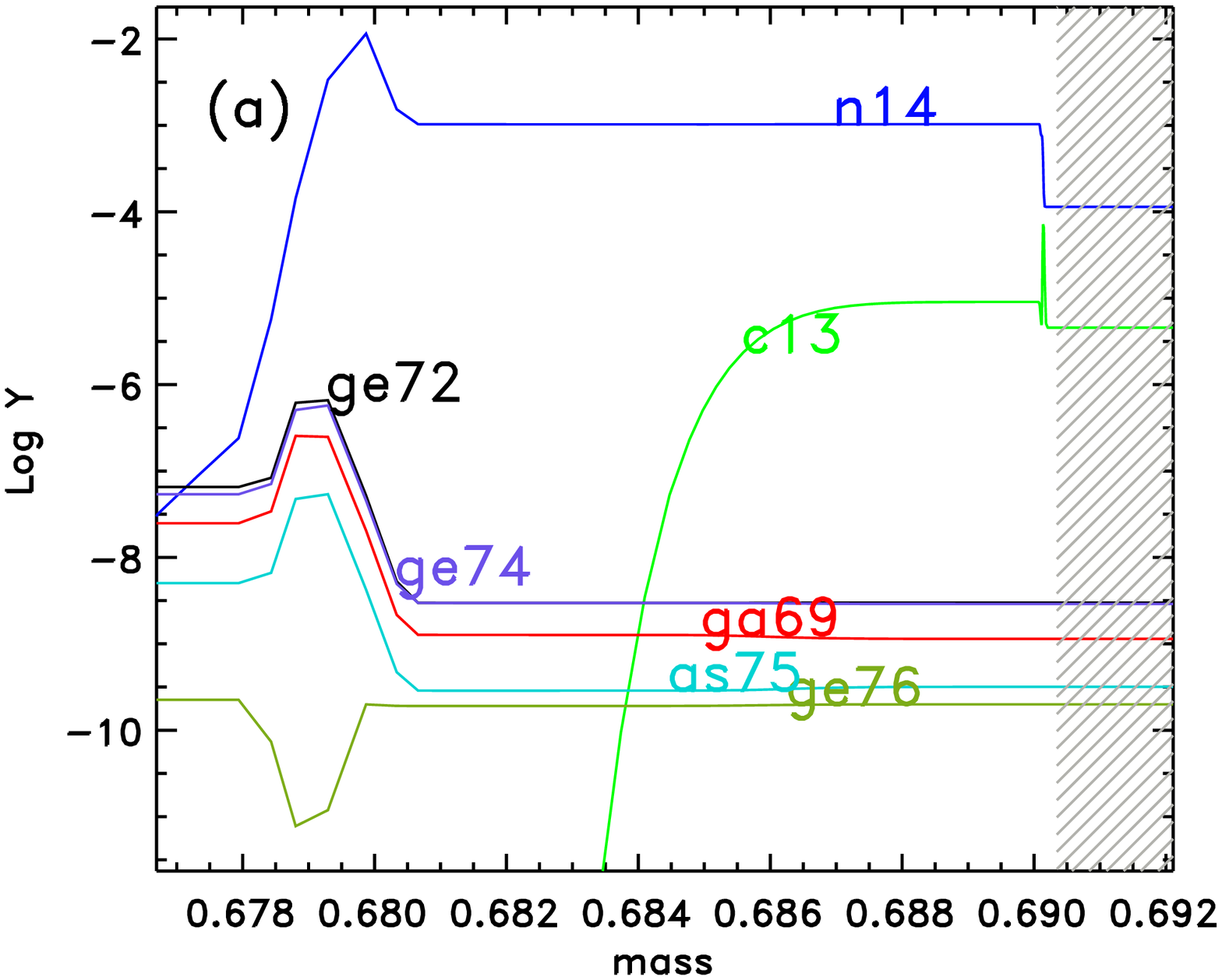} \\
\includegraphics[width=7cm,angle=90]{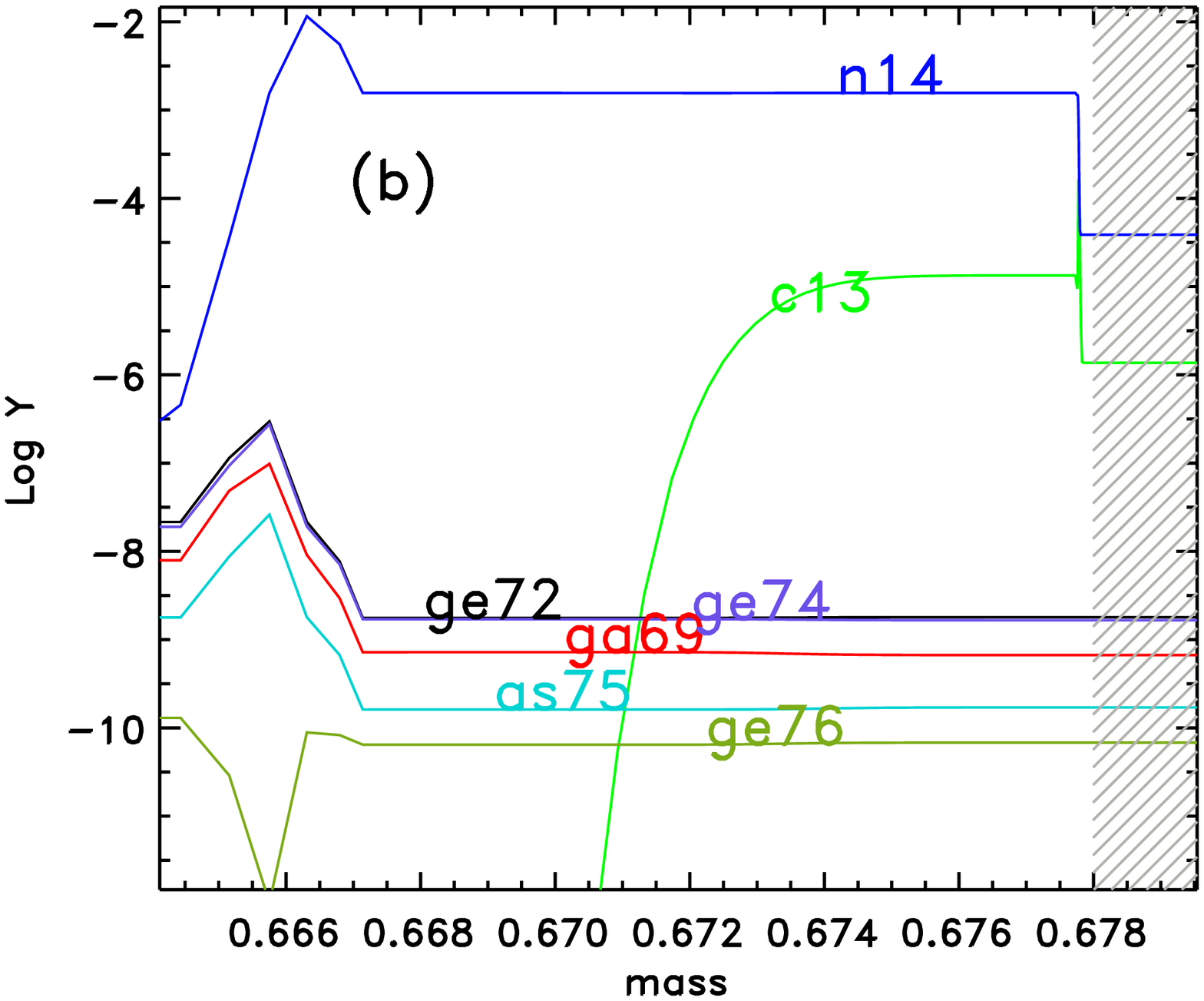}
\end{tabular}
\caption{Composition profile showing the intershell abundances 
(in $\log Y$, where mass fraction $X = Y A$) just before the last computed TP.
The shaded region is the inner edge of the convective envelope, and a PMZ of
$0.002\Msun$ was used in both models.  We show abundances from (a) the 3$\Msun$, 
$Z=0.012$ model where the Ge intershell abundance after the last TP is 
40 Ge$_{\odot}$, and (b) the 2.5$\Msun$, $Z= 0.004$ model where the Ge intershell
abundance is 63 Ge$_{\odot}$.  Note that the intershell abundances are typically
diluted by one order of magnitude at the surface by the last TDU episode.
\label{intershell}}
\end{center}
\end{figure}

In Table~\ref{results} we show the surface [Ge/Fe] and [Ga/Fe] ratios 
after the last computed TP for each stellar model, where we use Fe as 
the reference element which is justified because the surface abundance is
predicted by the models to be unaltered by AGB 
nucleosynthesis\footnote{Sulfur is similarly not altered.}. We also
provide the surface [Ge/S] and [Ge/S] ratios for the 3$\Msun$, $Z = 0.012$ 
model (with a PMZ $= 2\times 10^{-3}\Msun$)
for comparison. There is no difference between these values and 
those computed using Fe as the reference element.
\citet{sterling03} tentatively detected Ga in one 
PN (SwSt 1), with the derived [Ga/S] = 1.64 -- 2.87, depending on the 
level of depletion into dust. Table~\ref{results} shows that the largest 
predicted enhancements in terms of [Ge/Fe] were $\lesssim 0.56$ 
(or a factor of $\sim 3.6$) for all cases but the 2.5$\Msun$, $Z=0.004$ 
model, where [Ge/Fe] = 0.77.  Models with no \iso{13}C pocket produced very 
little Ge and Ga, except in the case of the 6.5$\Msun$ solar metallicity 
models.  For these massive AGB models, neutrons are released by efficient 
activation of the \iso{22}Ne neutron source during TPs and the increase 
of Ge at the tip of the TP-AGB is a factor of $\sim\,$1.5 above solar.
In comparison the 5$\Msun$, $Z=0.02$ model produced little Ge with 
([Ge/Fe] = 0.112) or without (0.04) a PMZ.

The 2.5$\Msun$, $Z=0.004$ model produced the most Ge, partly 
because efficient TDU results in a large amount of matter dredged from the 
core into the envelope (see Table~\ref{models}).
Another reason is that the efficiency of the \iso{13}C($\alpha,n$)\iso{16}O 
neutron source does not depend on the initial $Z$, and that means more
neutrons per Fe-seed nuclei are produced at lower metallicity. 
Table~1 from \citet{goriely00} also shows more Ge in the dredged-up 
matter with decreasing $Z$. \citet{gallino98} Figures~18 and 20
show the $s$-process enhancement factors in the intershell
material cumulatively dredged into the envelope in 2$\Msun$ models with 
$Z=0.01$ and $Z=0.0005$, respectively. The lower $Z$ model
clearly shows an increase in the production of the Ge isotopes,
with atomic mass $\approx 70$.

\begin{table}
\begin{center}
\caption{[Ge/Fe] and [Ga/Fe] ratios after the last computed TP\tablenotemark{a}. 
All masses are in solar masses.
\label{results}}
\vspace{1mm}
\begin{tabular}{@{}cccccc@{}}  \hline \hline
 Mass &  $Z$   & $M_{\rm csh}$ & PMZ & [Ge/Fe] & [Ga/Fe] \\ \hline
 3.0  &  0.02  & 1.6($-2$) & 2.0($-3$) & 0.474 & 0.453 \\
 5.0  &  0.02  & 4.9$(-3)$ & 1.0($-4$) & 0.112 & 0.111 \\
 6.5  &  0.02  & 2.3($-$3) & none & 0.191 & 0.212 \\
 3.0  &  0.012 & 1.6$(-2)$ & 1.0($-3$) & 0.386 & 0.482 \\
      &        &           & 2.0($-3$) & 0.557 & 0.678 \\
      &        &           &   ``  & 0.557\tablenotemark{b} & 0.677\tablenotemark{b} \\
 6.5  &  0.012 & 2.5$(-3$) & none & 0.186 & 0.304 \\
 1.9  &  0.008 & 2.0$(-2)$ & 2.0($-3$) & 0.252 & 0.333 \\
      &        &           & 3.0($-3$) & 0.339 & 0.439 \\
 2.5  &  0.004 & 1.7$(-2)$ & 2.0($-3$) & 0.777 & 0.894 \\
\end{tabular}

\tablenotetext{a}{Models with no PMZ have [Ge/Fe] $\approx 0$ for all cases but the
6.5$\Msun$ models.}
\tablenotetext{b}{These values are the [Ge/S] and [Ga/S] ratios, computed using
sulfur as the reference element.}

\end{center}
\end{table}

Our results compare favorably with the 3 $\Msun$ $Z=0.02$ model computed with
the FRANEC evolutionary code \citep{straniero97} by \citet{gallino98} 
with their standard \iso{13}C pocket (hereafter: the {\it Torino} models), 
who finds the final [Ge/Fe], [Ga/Fe], [Zr/Fe] and [Ba/Fe] equal to 0.46, 0.43, 1.03
and 0.93, respectively. The 1.5$\Msun$, $Z=0.02$ produced values equal to
0.28, 0.29, 0.75 and 0.65, respectively. 
The {\it Torino} 5$\Msun$ $Z=0.02$ model produced a final [Ge/Fe]=0.57, higher 
than our equivalent model, owing to the choice of the \citet{reimers75} mass 
loss (with $\eta = 10$) which results in more TPs (48) compared to our 
computation (24).  Massive AGB stars have been suggested as the progenitors 
of Type I bipolar PNe owing to their high He and N/O abundances 
\citep{stanghellini06}, as well as kinematics \citep{corradi95}.
\citet{sterling05c} and \citet{sterlingThesis} find a correlation between PNe
morphology and s-process overabundances, where elliptical PNe are more
enriched than bipolar PNe, which show little or no enhancement.
These observations suggest that Reimers type mass loss is inadequate for
modeling massive AGB stars, and formulae with a superwind 
\citep[such as][]{vw93} are favored.

None of the AGB models lost all of their envelopes
before the computation finished. That is, all of the models
were near the tip of the AGB but had not proceeded to the 
post-AGB phase. Some of these models could in principle  
experience further TPs and TDUs, and we refer to these
as {\em extra TPs}.  The amount of Ge expected 
after the $i^{\rm th}$ TP can be estimated according to 
$X^{i} = (M_{\rm env}^{i-1}X^{i-1} + \Delta M_{\rm Ge})/ M_{\rm env}^{i}$,
where $\Delta M_{\rm Ge} = \lambda \Delta M_{\rm intershell}$ 
is the mass of Ge dredged into the envelope, $\lambda$ is the TDU
efficiency and $\Delta M_{\rm intershell}$ is the mass of Ge in the
intershell.  In Figure~\ref{intershell} we show the intershell 
abundances for selected Ge isotopes 
for two models. A Ge pocket can be seen in both cases, and 
although this will be diluted by one order of magnitude by 
intershell convection, the post-pulse abundances are still 
enhanced compared to solar.

Using the 3$\Msun$, $Z = 0.012$ model with a PMZ of $2\times 10^{-3}\Msun$ 
as an example, we estimate that the next TP would take place when 
$M_{\rm env} \approx 0.7\Msun$. The mass of the He-intershell 
at the last TP is $0.016\Msun$, and assuming the same TDU
efficiency as in the previous TP, $\lambda = 0.835$. 
At the last computed time-step, the Ge envelope and intershell 
compositions are $9.64 \times 10^{-9}$ and $1.068 \times 10^{-7}$ 
(in mass fractions) respectively, from which we estimate that
after the next TP [Ge/Fe] = 0.64.  If we assume a mass-loss 
rate of $10^{-5} \Msun {\rm yr}^{-1}$ 
(typical near the end of the AGB) multiplied by the 
interpulse period of 54,000 years then 
$\sim 0.54\Msun$ is lost per interpulse\footnote{The same mass 
model computed with an interpulse period of 25,000 years near 
the end of the AGB would allow for 2 -- 3 more TPs.}.
From this we estimate that the final TP would occur when
$M_{\rm env} \sim 0.15\Msun$.  We can speculate that if a TDU
episode occurs with the same efficiency as before,
and using the envelope and intershell Ge abundances from 
the previous estimate ($1.17 \times 10^{-8}$), we get [Ge/Fe] = 0.90 
(or Ge $= 2.14 \times 10^{-8}$ in mass fractions), at the
tip of the AGB, close to the value required to match the 
composition of the most Ge-enriched PNe.
However caution is required because this value is very likely to be
an overestimate, owing to the evidence for a decreasing TDU 
efficiency with a decreasing envelope mass \citep{straniero97}.
Variations in AGB mass-loss rates and model-dependent
parameters such as the interpulse period will also effect this
estimate.

Increasing the size of the PMZ in the models produces more
Ge  but if we increase the size of the pocket too much we run 
into the problem that the PMZ accounts for a large percentage of 
the intershell region.  This is contrary to simulations which 
suggest that the \iso{13}C pocket should be at most $\sim$ 
10\% of the intershell \citep{denissenkov03,lugaro03,herwig05}.

\section{Discussion and conclusions}

The results presented in this Letter show that AGB models can explain
the spread of Ge abundance in PNe from solar to a factor of $\sim$ 10 
above solar, if we allow for the contribution of extra TPs not modeled
in detail. Most detailed models produced Ge overabundances up to a 
factor of $\sim 3$, corresponding to a dilution of 1 part of 
He-intershell material per roughly 20 parts of envelope material.
Hence, we can easily account for the Ge composition
of the majority of the observations, in particular taking into account
that substantial observational uncertainties arise from the choice of 
the level of dust depletion.  Even for the
mostly Ge-enriched PNe (SwSt~1 and BD $+$303639), for which Ge 
enhancements have been estimated to be up to a factor of 30, 
\citep[but could also be as low as $\simeq$ 5 if a lower dust depletion 
is assumed][]{sterling03,sterling05c},
it is conceivable that these overabundances could be 
reproduced by considering in more detail the effects of stellar model 
uncertainties, such as the mass loss and treatment of mixing processes which
affect the TDU in particular.

The relation between PNe with H-deficient CSs and those 
with the largest Ge abundances could be explained if a late TP during
the post-AGB phase (when the envelope mass decreases below $\sim 0.01\Msun$),
was responsible for both. These events, described by \citet{werner06}, 
cause the CS to become (partially) H-deficient by mixing and/or burning the small 
remaining envelope with the larger He-intershell, which is enriched in 
He-burning and s-process products. However such a scenario would also produce 
larger enhancements of the other $s$-process elements in this type of PNe,
while observations show that, on average, the enrichment of Kr and Se in PNe 
with H-deficient CS are not significantly different from other PNe 
\citep{sterling06,sterlingThesis}. \citet{zhang06} recently presented 
abundances for Br, Kr and Xe derived from PNe observations.
For all of these other elements, 
which we will discuss in detail in a forthcoming paper, results from the 
{\it Torino} code generally produce a good match with the observations 
(except in the case of Br which is not observed to be enhanced in PNe
contrary to model predictions), with higher enrichments obtained at 
lower metallicities, as in the case of Ge discussed here. 
On the other hand, tentative evidence 
for the late TP scenario could come from BD $+$303639, with enhanced Ge 
and an Fe-deficient CS \citep{werner06}, possibly indicating a high 
level of mixing of $s$-processed material. Further study of the stellar model
uncertainties and the extra and/or late TP scenario together with more
accurate observations are needed before anything more conclusive can be
said.


%% Included in this acknowledgments section are examples of the
%% AASTeX hypertext markup commands. Use \url without the optional [HREF]
%% argument when you want to print the url directly in the text. Otherwise,
%% use either \url or \anchor, with the HREF as the first argument and the
%% text to be printed in the second.

\acknowledgments

AIK warmly thanks Onno Pols at the Astronomical Institute at Utrecht
University for his hospitality during a visit while writing this paper.
ML gratefully acknowledge the support of NWO through a VENI grant. 
RG acknowledges support from the Italian MIUR-FIRB project 
``The astrophysical origin of the heavy elements beyond Fe''.
We thank Nick Sterling for helpful discussions, and the anonymous
referee for a thorough report.

\end{document}